\documentclass[12pt,english,american]{article}
\usepackage[T1]{fontenc}
\usepackage[latin1]{inputenc}
\usepackage{geometry}
\usepackage{graphicx}
\usepackage{setspace}
\makeatletter

\providecommand{\tabularnewline}{\\}

 \newenvironment{lyxlist}[1]
   {\begin{list}{}
     {\settowidth{\labelwidth}{#1}
      \setlength{\leftmargin}{\labelwidth}
      \addtolength{\leftmargin}{\labelsep}
      }}
   {\end{list}}

\usepackage{babel}
\makeatother
\begin{document}

\title{The photomultiplier tube testing facility for the Borexino experiment
at LNGS. }

\maketitle
\begin{center}{\large A.Brigatti}%
\footnote{{\large I.N.F.N. sez. di Milano, Via Celoria, 16 I-20133 Milano, Italy}%
}{\large , A.Ianni}%
\footnote{I.N.F.N. Laboratorio Nazionale del Gran Sasso, SS 17 bis Km 18+910,
I-67010 Assergi(AQ), Italy%
}{\large , P.Lombardi}%
\footnote{Dipartimento di Fisica Universitá and {\large I.N.F.N. sez. di Milano,
Via Celoria, 16 I-20133 Milano, Italy}%
}{\large , G.Ranucci$^{3}$,  and O.Smirnov}%
\footnote{Corresponding author: Joint Institute for Nuclear Research, 141980
Dubna, Russia. E-mail: osmirnov@jinr.ru;smirnov@lngs.infn.it%
}\end{center}{\large \par}

\begin{abstract}
A facility to test the photomultiplier tubes (PMTs) for the solar
neutrino detector Borexino was built at the Gran Sasso laboratory.
Using the facility 2200 PMTs with optimal characteristics were selected
from the 2350 delivered from the manufacturer.

The details of the hardware used are presented. 
\end{abstract}

\newpage
\section{Introduction}

\subsection{The Borexino experiment}

Borexino, a large scale liquid scintillator detector under construction
at the Gran Sasso underground laboratory (LNGS), is a low energy solar
neutrino detector designed for $^{7}Be$-neutrinos registering\cite{Borex}.
The scintillation photons produced by the recoil electrons are detected
by 2200 photoelectron multiplier tubes (PMTs) placed around a transparent
inner vessel containing a scintillating mixture. In addition 200 PMTs
are used in the external Cherenkov light muon veto system. The Borexino
design is described in detail elsewhere \cite{Borex2}.

The Monte Carlo simulation of the Borexino detector predict that the
mean number of photoelectrons (p.e.) registered by one PMT will be
in the region $0.02-2.0$ for an event with energy of 250-800 keV.
The interaction point in the detector is reconstructed using the time
information from PMTs. The position resolution depends therefor on
the precision of measurement of the arrival time of a single photoelectron.
The accurate energy measurement in its turn demands good single electron
charge resolution. Hence, the PMTs should demonstrate a good single
electron performance both for the amplitude and the timing response.
Furthermore, in order to minimize the probability of the random triggers
during the acquisition, the PMT should feature a low dark rate. Another
parameter to be kept under control is the probability of the delayed
trigger of the system which depends on the PMT afterpulsing rate (mainly
ionic afterpulses). The overall required detector performances define
the acceptable characteristics for the PMTs, which are summarized
in table \ref{AcceptanceChar}. After preliminary tests (\cite{Philips,Hamamatsu,R93,RG93}),
the ETL9351%
\footnote{at the time of the tests this type of photomultiplier was named Thorn
EMI9351%
} with a large area photocathode (8'') has been chosen \cite{ETL}.
Because of very low radioactive background needed for the operation
of the Borexino detector, very strict limits were set on the content
of the radioactive impurities in the PMT constructive materials. All
the PMT components were thoroughly examined for the content of the
radioactive elements from the $^{238}U$ (design radiopurity of $<3\times10^{-8}$
g/g) and $^{232}Th$ chains ($<1\times10^{-8}$ g/g), and $^{40}K$
($<2\times10^{-5}$ g/g of natural $K$ content) in the glass of the
PMT. The radioactivity measurements were performed with $Ge$ spectrometry
at the Gran Sasso underground laboratory. The design radiopurities
of all the PMTs components were achieved, the detailed report on the
measurements can be found in \cite{Radioactivity}. 

Before installation all PMTs have been tested in the special facility.
The procedure of the testing includes adjustment of the operating
voltage in order to set the multiplier gain to $k=2\times10^{7}$. 

In August 2001 bulk testing of 2350 PMTs for the Borexino was completed,
and 2000 PMTs with the best performances were selected for the installation
in the Borexino detector. The high efficiency of the equipment permitted
to complete PMT testing within 4 months. 

\begin{table*}
\begin{center}\begin{tabular}{|c|c|c|c|c|c|c|}
\hline 
Parameter&
$f_{dark}$&
$p/V$&
$\sigma_{t}$&
$p_{late}$&
$p_{prep}$&
$p_{a}$\tabularnewline
\hline 
&
<20 Kcps&
>1.25&
<1.25 ns&
<5\%&
<1\%&
<5\%\tabularnewline
\hline
\end{tabular}\end{center}

\caption{\label{AcceptanceChar}Acceptance characteristics of a PMT:}

\begin{lyxlist}{00.00.0000}
\item [$f_{dark}$]dark rate (with a discriminator threshold set at 0.2
p.e. level);
\item [$p/V$]peak-to-valley ratio (the ratio of the peak value in the
charge spectrum to the value at the valley between the low- amplitude
pulses and the main peak);
\item [\textbf{$\sigma_{t}$}]the rms of the Gaussian fitting the main
peak in the transit time distribution;
\item [$p_{late}$]late pulsing in percent, estimated as the ratio of the
events in the $[t_{0}+3\cdot\sigma_{t};100]$ ns range to the total
number of the events registered in the 100 ns interval;
\item [$p_{prep}$]prepulsing in percent, estimated as the ratio of the
events in the $[\textrm{t}_{0}-20;t_{0}-3\cdot\sigma_{t}]$ ns range
to the total number of the events;
\item [$p_{a}$]total amount of afterpulses following the single electron
response in the time interval up to 12 $\mu s$ (measured in percent
in respect to the amount of the main pulses).\end{lyxlist}

\end{table*}

\subsection{8'' PMT ETL9351 and its characteristics}

\begin{figure*}
\begin{center}\includegraphics[%
  width=0.60\textwidth,
  height=0.90\textwidth]{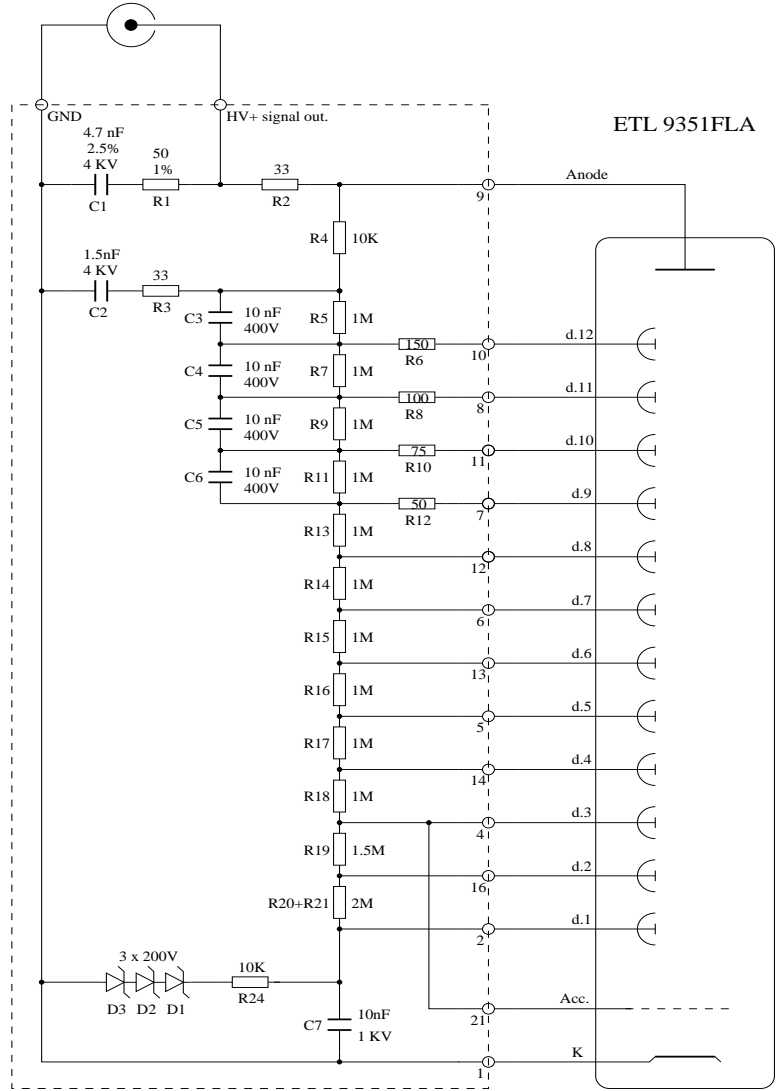}\end{center}

\caption{\label{fig:divider}The base used for the Borexino PMTs.}
\end{figure*}

A PMT of this model has 12 dynodes with a total gain of $k=10^{7}$.
The transit time spread of the single p.e. response is about $1$
ns. The PMT has a good energy resolution characterized by the peak-to-valley
ratio. The manufacturer (Electron Tubes Limited ETL) guarantees a
peak-to-valley ratio of 1.5. 

\begin{figure*}
\begin{center}\includegraphics[%
  width=0.80\textwidth,
  height=0.25\textwidth]{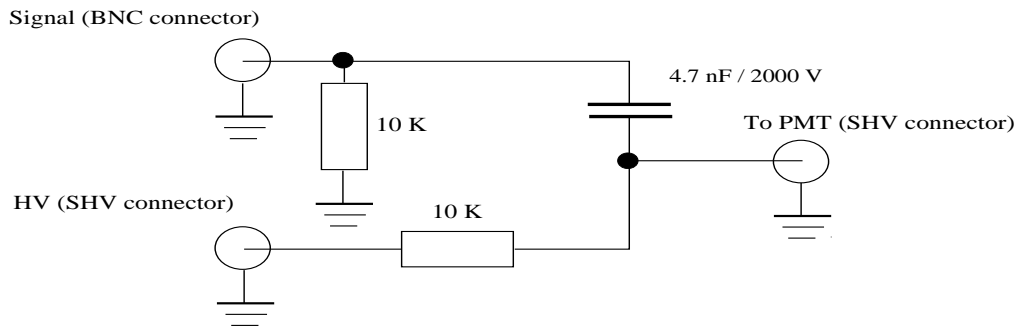}\end{center}

\caption{\label{Decoupler}Signal decoupler used with the Borexino PMTs}
\end{figure*}

The PMTs delivered are factory tested and the operational high voltage
(HV) is specified by the manufacturer. Selective measurements of the
PMTs at the specified voltage showed a high variance of the gain near
its mean value of $k=(0.86\pm0.25)\times10^{7}$. Moreover, the high
voltage divider used by ETL and the one used in Borexino are different.
For these reasons the operational voltage was readjusted. The Borexino
PMT divider is shown in Fig.\ref{fig:divider}. The signal and the
high voltage are carried on the same cable. The signal decoupler scheme
is shown in Fig.\ref{Decoupler}. For proper signal termination a
50$\Omega$ resistor R1 is included in the anode chain. The resistor
decreases the signal amplitude by a factor of 2. The signal attenuation
is compensated by the higher operating voltage. Thus actually, the
high voltage is adjusted for each PMT in order to provide photoelectron
multiplier gain $k=2\times10^{7}$. 

\section{The PMT test facility at LNGS}

Several PMT test facilities were prepared in the past by collaborations
planning to use big number of PMTs in their detectors, such as the
IMB \cite{IMB}, CHOOZ \cite{CHOOZ} and SNO \cite{SNO} experiments.
Also for the Borexino experiment a special PMT test facility was prepared
at the LNGS. The photomultipliers used in the prototype of the Borexino
detector, the Counting Test Facility (CTF) \cite{CTF}, were tested
with a simpler facility, described in \cite{RG93}. The Borexino PMT
test facility is placed in two adjacent rooms. In one room the electronics
is mounted, while the other one is a dark room with 4 wooden tables
designed to hold up to 64 PMTs. The tables are separated from each
other by black shrouds, which shield the light reflected from the
PMTs photocathode. Together with the acceptance tests, performance
of the PMTs immersed in water was tested, in order to study the possible
mechanical problems for PMTs to be mounted at the bottom part of the
Borexino detector. For this purpose 3 pressurized Water Tanks (WT)
were installed, each of the tanks can hold up to 20 PMTs. For the
purpose of the long-term testing of the PMTs in conditions close to
the ones of the Borexino detector a Two-Liquid Test Tank (TLTT) was
designed and installed in the laboratory \cite{Fibers}. 48 PMTs are
installed in the TLTT, with the base being in water and the bulb immersed
in liquid scintillator. Actually, the TLTT test has now been running
for more than 3 years, data is taken periodically in order to check
the PMT signal degradation. Only two PMTs of the total 48 are found
non-operational after these 3 years.

The overall test facility is therefore able to measure the characteristics
of 172 PMTs if completely loaded. The electronics system consists
of the high voltage supplies and the independent signal processing
electronics. The high voltage supplies can provide high voltage to
all PMT under test. In order to avoid unnecessary system duplication,
only 32 electronics channels are used for the signal processing at
the same time. The PMTs under test are grouped by no more than 32
in one group. The switching between the groups is performed by means
of 8 analogical multiplexers, which provide control over 4 complete
groups. Two groups of 32 PMTs each are connected to the 64 cables
of the dark room. Another two groups can be connected interchanging
the cables either to the 60 PMTs in the WTs (20 PMTs each), or to
the 48 PMTs of the TLTT (24 PMT in each group). The tests with the
TLTT are rare and the connection of the signal cables is performed
manually. 

The dark room is equipped with a system compensating the Earth's magnetic
field (\cite{EMF}). The non-uniformity of the compensated field in
the plane of the tables is no more than $10\%$. In the WTs and the
TLTT the Earth Magnetic Field is not compensated.

The PMT characteristics are defined during a 8 hours run. The stability
of the parameters is defined every 12-24 hours during longer runs.

\subsection{The Earth magnetic field compensation system}

Magnetic fields as weak as the Earth's magnetic field (EMF, $\sim40\mu T$)
affect PMT performances and it turns out that tubes with linear focusing
dynodes (as the Borexino 8\char`\"{} ETL9351) are most sensitive to
magnetic influence when the field is parallel to the dynodes~\cite{R93,caputa96}.

In the area of the external Gran Sasso Laboratory the static EMF is
about 35$\mu$T in the vertical direction, 25$\mu$T along the north-south
direction and 8$\mu$T along the east-west direction. A daily change
around 0.05$\mu$T was observed. Therefore, a compensation of the
north-south and vertical components of the static EMF was needed;
the component varying in time is negligible and can be left without
compensation. Since the test facility tables occupy in total 2x2 m$^{2}$,
a large compensated volume is needed. The EMF compensation system
of the test facility is based on a system of rectangular coils%
\footnote{Many coils systems have been studied in order to produce an homogeneous
magnetic field~\cite{sidney45,merritt83,grisenti81}.%
}; the choice is characterized by simple construction, easy access
to the internal area, and by an excellent ratio of uniform field to
coils volume~\cite{merritt83}. Since the dark room is in iron-reinforced
concrete building with dimensions of 5x5x3 m$^{3}$ made of two sandwich-walls,
one should expect spatial variations of the EMF over the useful volume.
Therefore, the EMF was mapped inside the dark room. Moreover, in order
to avoid any interference with the compensation system, the support
structure for the phototubes has been manufactured with wood.

A number of various systems of rectangular and square coils was studied.
A four square coils system (FCMS) as designed by R. Merritt and coworker~\cite{merritt83}
was chosen in order to compensate the crucial north-south component.
In order to analyze the magnetic field created by the FCMS a master
formula for a general rectangular coil was written as described in
\cite{EMF}. The field uniformity inside the coils system, $U$, was
characterized by the field deviations with respect to the field at
the center: \begin{equation}
U=\arrowvert\frac{\overrightarrow{H}(x,y,z)-\overrightarrow{H}(0,0,0)}{\overrightarrow{H}(x,y,z)}\arrowvert.\label{EMF:Eq1}\end{equation}
 \foreignlanguage{english}{}where $\overrightarrow{H}(x,y,z)$ is
the field at the point $(x,y,z)$ and $\overrightarrow{H}(0,0,0)$
is the field at the center of the system. The field $\overrightarrow{H}(0,0,0)$
was chosen to fully compensate the measured EMF at the dark room center.
The coil currents were calculated on the basis of this value.

A FCMS with 2.84m coils side length was chosen to compensate the north-south
component. Another system consisting of two square coils with 2.98~m
side length and spacing equal to 1.4m%
\footnote{The Helmholtz separation for such a system is 1.62m~\cite{cacak69}.
This spacing is not the best solution to achieve the maximum uniformity
over a large volume with two square coils~\cite{ference40}.%
} was chosen to compensate the vertical component of the EMF. The effect
of the finite cross section of the coils (66 mm$^{2}$) was taken
into account as described in~\cite{ference40}, but the corrections
were found to be negligible because of the huge coil sizes.

\begin{figure*}
\begin{center}\includegraphics[%
  width=1.0\textwidth,
  height=0.90\textwidth]{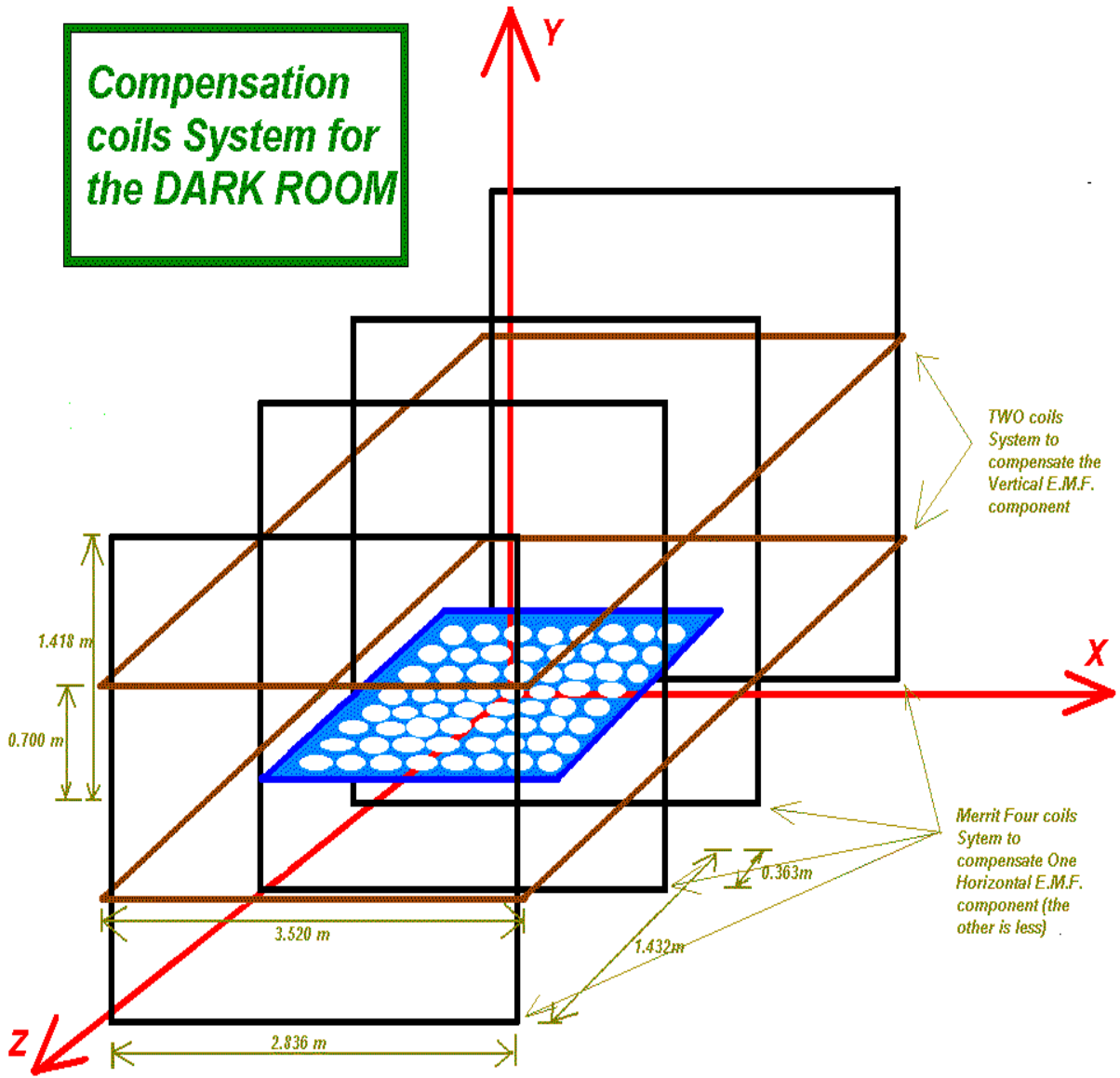}\end{center}

\caption{\label{coilssystem}Coil system and reference frame.}
\end{figure*}

The useful volume for arranging the PMTs was defined determining the
10\% deviation of the magnetic field from the mean value in a plane
crossing the center of the coil system. All field components were
treated independently. In Fig.\ref{coilssystem} is shown the coil
system with it reference frame. The measurements confirmed a good
uniformity of the EMF in a volume of 2x2x0.6 m$^{3}$ lying in the
xz plane. In Table~\ref{tabcoils} all the parameters of the six
coil system are reported.

\begin{table}

\caption{Coil system parameters. \label{tabcoils}}

\begin{center}\begin{tabular}{|l|c|c|}
\hline 
\multicolumn{3}{|c|}{Four coil system (FCMS) }\tabularnewline
\hline
&
 inner coils &
 outer coils \tabularnewline
\hline
side length &
 2.84m &
 2.84m \tabularnewline
\hline
distance to the center &
 0.363m &
 1.434m \tabularnewline
\hline
current &
 0.34 A &
 0.80 A \tabularnewline
\hline
number of turns &
 47 &
 47  \tabularnewline
\hline
\end{tabular}\end{center}

\begin{center}\begin{tabular}{|l|c|}
\hline 
\multicolumn{2}{|c|}{Two coils system (square Helmholtz coils)}\tabularnewline
\hline
side length &
 2.98m \tabularnewline
\hline
distance between the coils&
 1.4m \tabularnewline
\hline
current &
 1.11 A \tabularnewline
\hline
number of turns &
 50  \tabularnewline
\hline
\end{tabular}\end{center}
\end{table}

The measurements of the EMF inside the dark room were performed with
an analog Magnetoscope 1.608 by Foerster which is an intensity and
gradient magnetic field sensor operating with a Foerster probe. The
precision of the instrument is 3\% for all the scales. Two sets of
measurement were performed at the level of 146~and 129~cm from the
ground. The results of measurement were reported in \cite{EMFcompensation}.
On the basis of the analysis of the field uniformity, the PMT support
tables were placed in a suitable area (far away from the entrance
door and the cables patch-panel). Measurements of the effect of movements
of a big bridge crane located in vicinity of the dark room were performed
as well~\cite{EMF}.

\begin{figure*}
\begin{center}\includegraphics[%
  width=0.90\textwidth,
  height=0.90\textwidth]{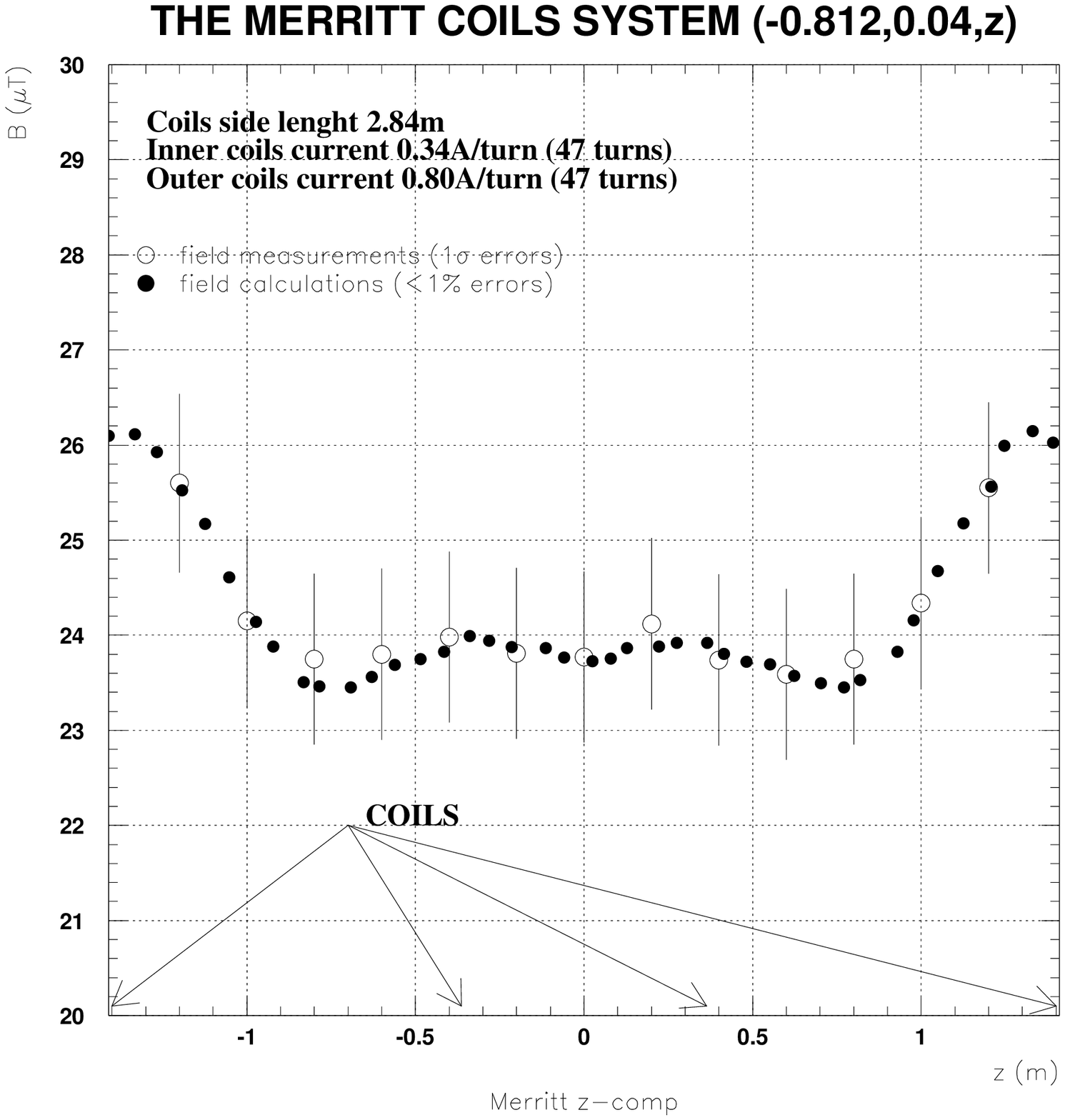}\end{center}

\caption{\label{merrits}Magnetic field measurements. See text for details.}
\end{figure*}

\begin{figure*}
\begin{center}\includegraphics[%
  width=0.90\textwidth,
  height=0.90\textwidth]{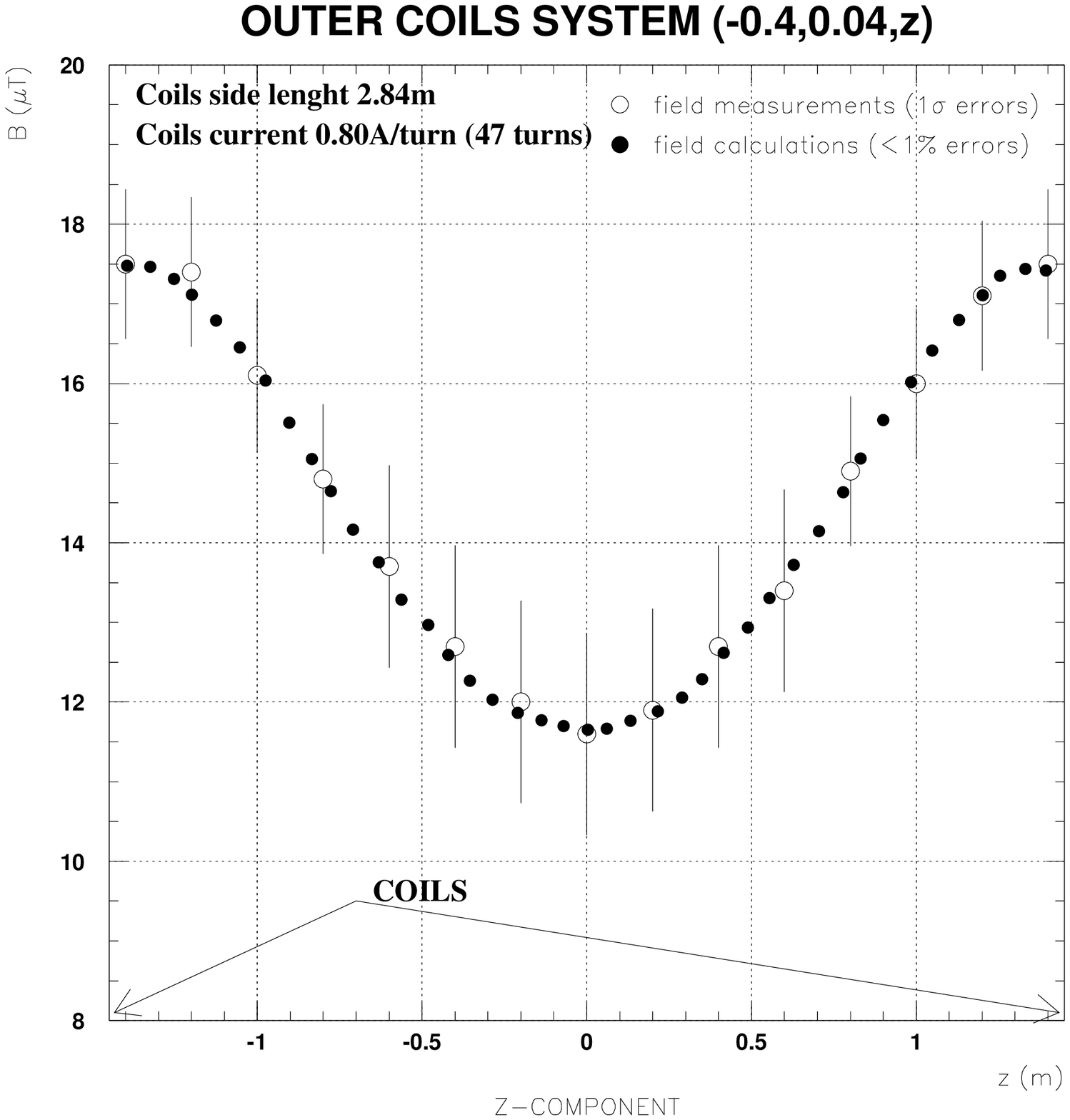}\end{center}

\caption{\label{outer}Magnetic field measurements. See text for details.}
\end{figure*}

The power supply for the 6 coil system inside the dark room (see Fig.\ref{coilssystem})
is guaranteed by a single double Elind generator which is able to
provide a stabilized current up to 0.01~A. Each coil consists of
a wooden structure with three layers of copper-wire turns (see Table~\ref{tabcoils}).
Fig.\ref{merrits} and Fig.\ref{outer} present the results of the
final measurements after the coil installation.

\begin{figure*}
\begin{center}\includegraphics[%
  width=0.90\textwidth,
  height=0.90\textwidth]{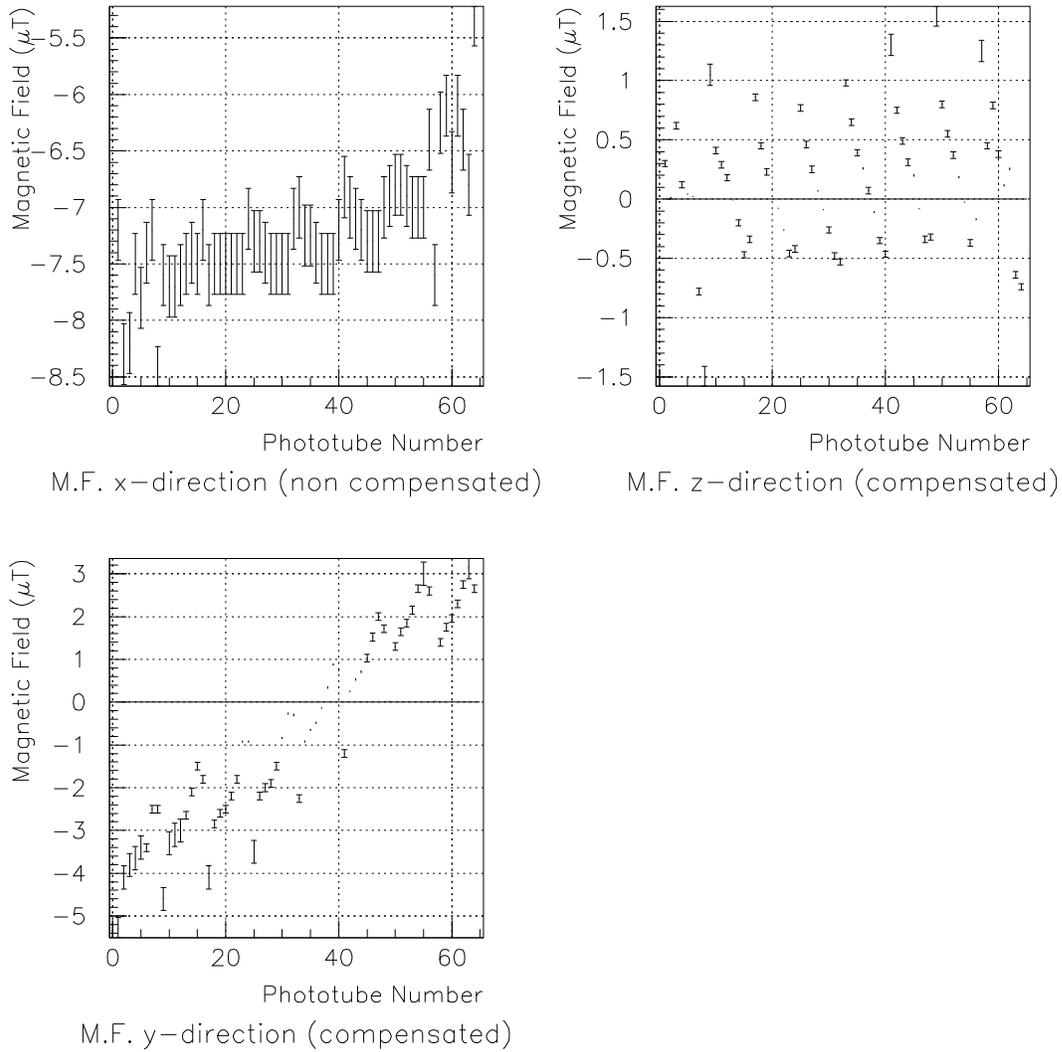}\end{center}

\caption{\label{holes}Residual field measured at the center of each PMT position
inside the dark room.}
\end{figure*}

Once the coil system was built and operating the measurements were
performed in order to check the compensated field. The tables for
the PMTs have in total 64 holes. The field at the center of each hole
has been measured. Fig.\ref{holes} shows the results of measurements
with the compensated field. One can see that the developed coils system
compensates the EMF at the level of 1.5~$\mu$T for the north-south
component (FCMS), and 5~$\mu$T for the vertical component (non standard
Helmholtz system of square coils). The value of 8.5~$\mu$T is measured
for the non-compensated east-west component. The expected PMT signal
degradation due to the residual EMF is less than 5\%~\cite{R93}. 

\begin{figure*}
\begin{center}\includegraphics[%
  width=0.90\textwidth,
  height=0.40\textwidth]{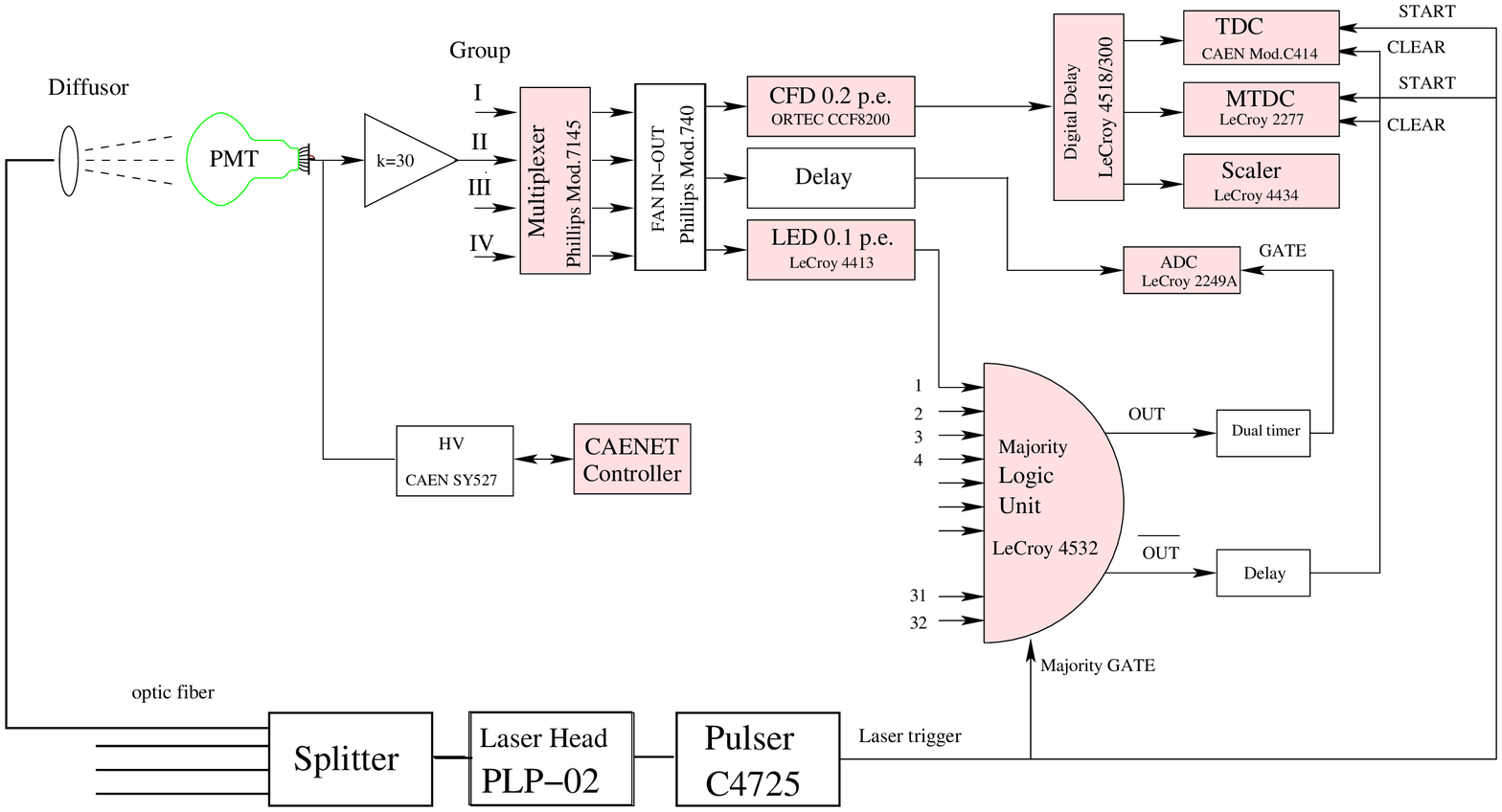}\end{center}

\caption{\label{fig:electronics}One channel of the electronics used in the
measurements; a PMT of Group II is shown. CAMAC standard modules controlled
(and/or read) by a personal computer via a CAEN C111 interface are
shown in gray color. }
\end{figure*}

\subsection{Electronics}

One channel of electronics (out of the total 32) of the test facility
is presented in Fig.\ref{fig:electronics}. The system uses the modular
CAMAC standard electronics and is connected to a personal computer
by a CAEN C111 interface. 

\begin{figure*}
\begin{center}\includegraphics[%
  width=0.80\textwidth,
  height=0.25\textwidth]{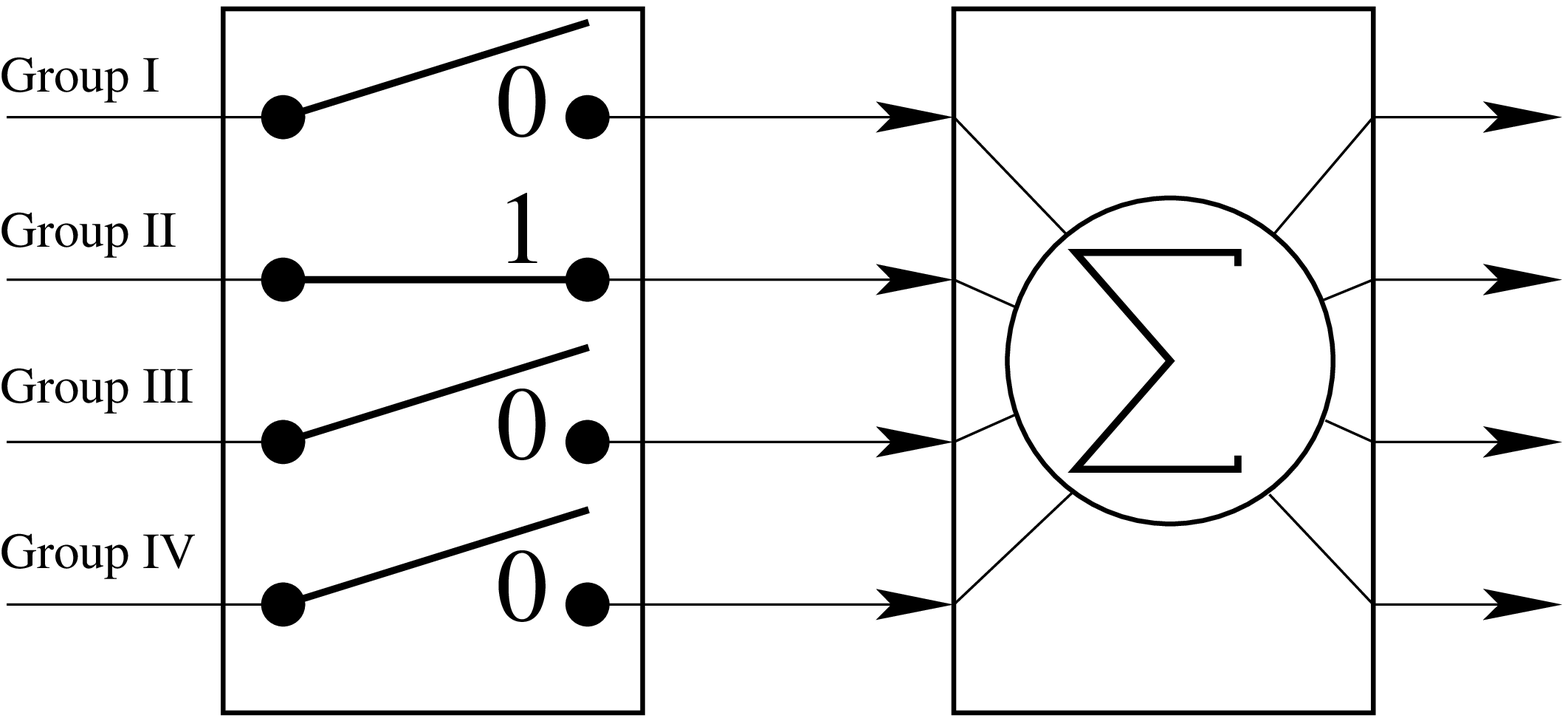}\end{center}

\caption{\label{Figure:MUX}The schematic of the channel multiplexing}
\end{figure*}

As it was already pointed out, the test system operates only on one
of 4 groups of PMTs at the same time. The logic of the channels multiplexing
is shown in Fig.\ref{Figure:MUX}. The switch between the groups is
performed writing the corresponding mask into the multiplexer internal
register. The Quad linear FAN-IN/OUT (Philips Mod.740) is connected
as analog signals summator. The signal is amplified before multiplexing
in order to preserve the signal-to-noise ratio. After splitting the
signals are processed by a LeCroy Mod.4413 Leading Edge Discriminator
(LED) with a threshold set at the level of $0.05-0.10$ p.e., just
above the level of the noise. If the PMT in the current channel is
too noisy, the channel is disabled by writing the corresponding mask
into the internal register of the LED, in such a way avoiding the
influence of the elevated count rate in one channel on others. The
signals formed by the LED are fed into the LeCroy Mod.4532 Majority
Logic Unit (MALU). If at least one signal from the total 32 occurs
inside the external gate, the output signal is generated and the LAM
signal is activated by the MALU. The laser trigger is used as the
majority external gate.

The Ortec Mod.CCF8200 Constant Fraction Discriminator (CCF) was used
for the measurements of the timing characteristics with the threshold
set at the 0.2 p.e. level. The formed signals are delayed and split
on the Digital Delay (LeCroy Mod.4518/300) and then fed into the CAEN
Mod.C414 Time-to-Digital-Converter (TDC), the LeCroy Mod.2277 Multihit
TDC (MTDC) and the Scaler (LeCroy Mod.4434). The original signal delayed
on the Delay Line is fed into the LeCroy Mod.2249A Analog-to-Digital
Converter (ADC). The ADC {}``gate'' and the TDC/MTDC {}``start''
signals are generated using the laser internal trigger, which has
a negligible time jitter ($<100\: ps$) with respect to the light
pulse. 

The MALU is able to memorize the pattern of the hit channels. This
information significantly increases the data processing rate. The
read-out of the electronics is activated when the majority LAM signal
is on. Otherwise, a hardware clear is forced 400 ns after the laser
trigger, using the majority $\overline{OUT}$ signal.

\begin{figure*}
\begin{center}\includegraphics[%
  width=0.90\textwidth,
  height=0.90\textwidth]{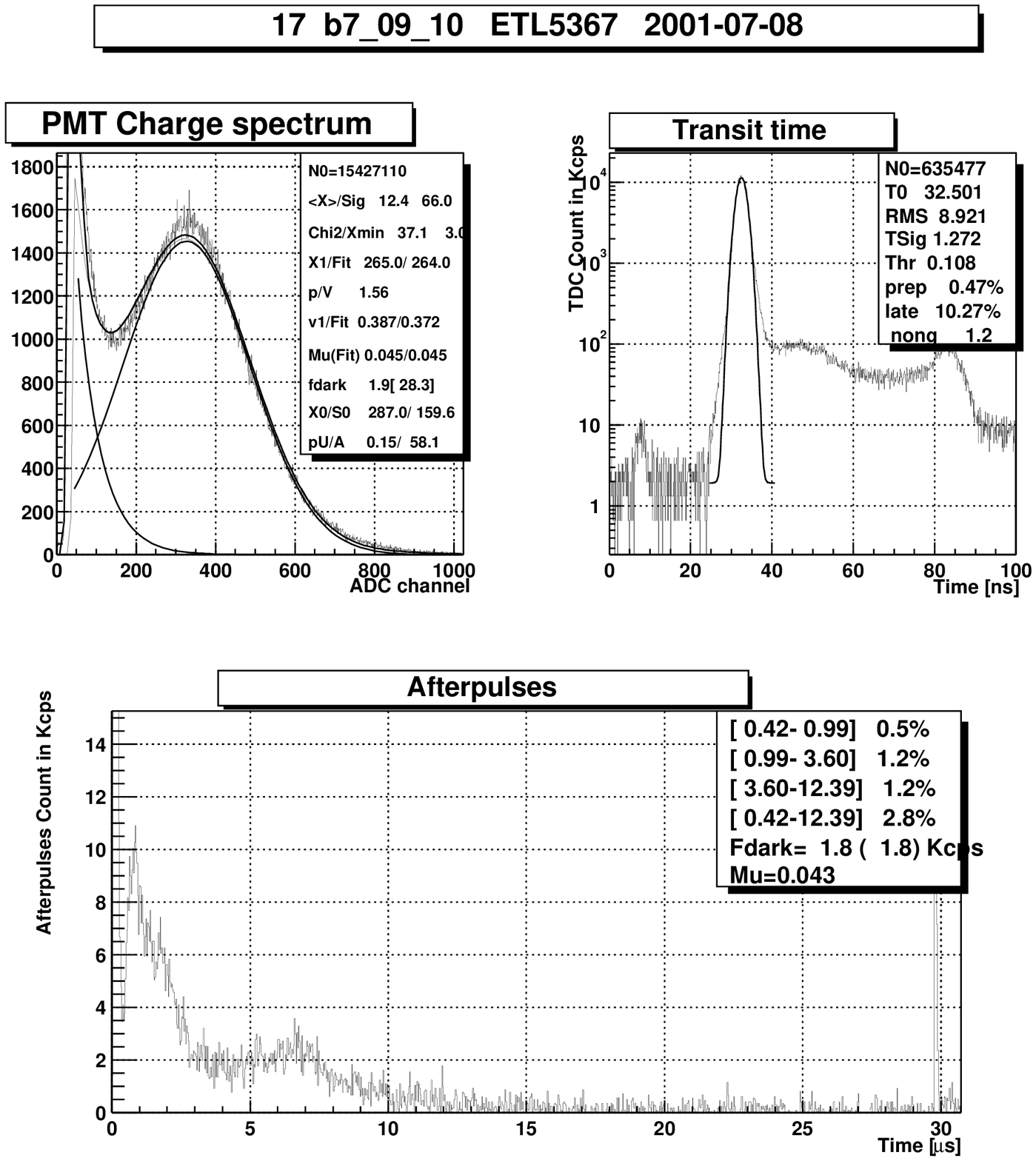}\end{center}

\caption{\label{fig:example}The characteristics of one of the PMTs under
test: charge spectrum, single photoelectron transit time and afterpulses.
The PMT charge response is measured in ADC channels (1024 ADC channels
correspond to 256 pC).}
\end{figure*}

Two modes of operation were used in our tests. In the first one, the
system is triggered at the first trigger from the laser that occurs
when the electronics is not busy processing the previous event. This
was realized by connecting the last (32-nd) majority input to the
external gate signal. This mode of operation has been used in the
test of the PMTs. An example of the data acquired during a routine
PMT test is presented in Fig.\ref{fig:example}. 

Another mode of operation has been used during the HV tuning. In this
case the 32-nd input of the MALU was disabled. The electronics read-out
was activated as before by the MALU LAM signal, but for this time
at least one signal from PMT over the LED threshold must be present
to activate the MALU. In such a way the {}``cut'' charge histograms
have been acquired with a hardware threshold of about $5\%-10\%$
of the {}``typical'' Single Photoelectron Response (SER) mean value. 

The afterpulses are registered by the MTDC which is able to record
in the internal register up to 16 sequential hits inside a 32 $\mu s$
window for each of its 32 channels. The laser repetition rate was
tuned to 30 $\mu$s, so the peak corresponding to this time can be
clearly seen at the afterpulses histogram. The intensity of this peak
was used to monitor the intensity of the laser. The probability of
two sequential nonzero PMT signals, in the assumption of the Poisson
distribution of the number of detected photoelectrons, is given by:

\[
\mu=-\ln(1-\frac{N_{2}}{N_{1}})\simeq\frac{N_{2}}{N_{1}},\]

for $\mu\ll1$, where $N_{2}$ is number of events in the second peak
and $N_{1}$- number of events in the first peak (or the total number
of the system triggers in our case). The time of arrival of the first
pulse was tuned to have it just after the MTDC start signal, so the
MTDC counts the number of its triggers.

The high precision calibration of each electronics channel was performed
before the measurements. Here calibration means the ADC response to
a signal corresponding to 1$\:$p.e.%
\footnote{multiplied by a factor of $2\times10^{7}$ by the electron multiplier
and providing $1.6\:$pC total collected charge at the PMT output%
} on the system input with the ADC pedestal subtracted (the PMT in
this measurement was substituted by a precision charge generator LeCroy
1976). The position of the ADC pedestals are defined and checked during
the run.

\subsection{High voltage supplies}

Two types of high voltage supplies were used to provide HV to the
PMTs. The CAEN mainframe mod. SY403, with 4 mod. A505 boards (3 kV,
200 $\mu$A) of 16 channels each, was used to provide HV for the PMTs
in the dark room. This model has very low r.f. noise and provides
the possibility to check the current on each channel, which was very
important during the first test of the PMTs.

The HV for the TLTT and three Water Tanks were provided by the Universal
Multichannel Power Supply System CAEN mod. SY527, where the mainframe
hosts 5 mod. A932AP boards (2.5 kV, 500 $\mu$A) of 24 channels each.
These modules have much higher r.f. noise and do not provide the possibility
of the read-out of the individual channel current; only the current
of the primary channel providing power for 24 channels can be checked.

The modules are remotely controlled through the CAEN mod. C117B H.S.
CAENET CAMAC Controller Interface and the H.S. CAENET serial link
and protocol. Two mainframes (SY403 and SY527) are connected in series.

\subsection{Reduction of the r.f. noise and stabilization of the system}

PMTs with a large area photocathode are very sensitive to environmental
radio-frequency noise. The PMTs tested for the CTF detector were sealed
in a plastic water-proof container; in order to reduce the induced
noise, the PMTs were wrapped in aluminum foil before testing. The
PMTs prepared for Borexino were sealed in a metallic cylindric container.
Tests showed the dependence of the noise amplitude on the width of
the gap between the internal aluminum screen and the edge of the metallic
cylinder. This noise had a smaller amplitude in comparison to the
non-protected PMTs of the CTF and no special measures were needed
in oder to operate the PMT at the acceptable level of the r.f. noise.

All the cabling was performed with care in order to avoid ground loops.
During the exploitation of the system another source of r.f. noise
was identified. This noise was originating from the flat cables, connecting
the personal computer with the CAMAC crate. In the following the cable
has been wrapped in aluminum foil, one end was grounded and the other
one left in the air.

The electronics room was equipped with an air conditioning system.
Nevertheless, a slow drift of the ADC pedestals was observed during
the long runs. In oder to compensate the slow variations of the pedestal
position, it was automatically checked every 30 minutes and in the
case a statistically significant drift was measured, a software compensation
was introduced.

\subsection{Optical system}

\begin{figure*}
\begin{center}\includegraphics[%
  width=0.35\textwidth,
  height=0.80\textwidth,
  keepaspectratio,
  angle=90]{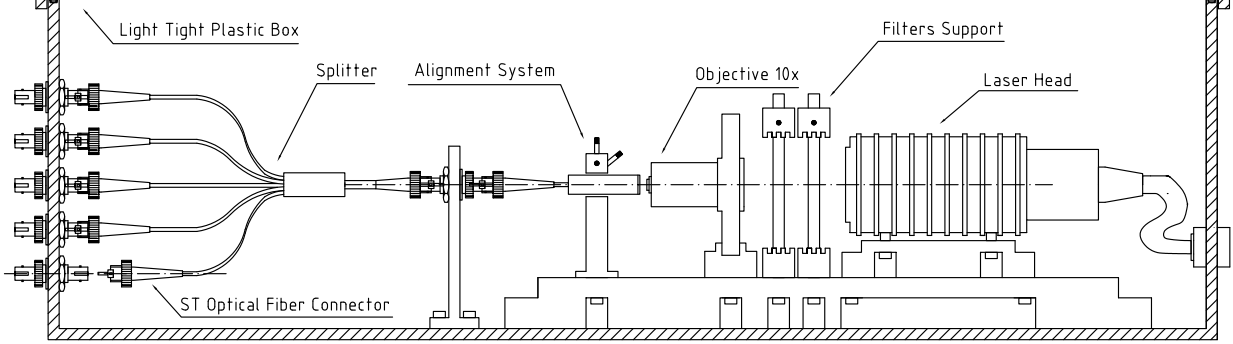}\end{center}

\caption{\label{Fig:OpticalSystem}Sketch of the optical system.}
\end{figure*}

The PMTs are illuminated by low intensity light pulses from a picosecond
Hamamatsu Photonics K.K. pulse laser, model PLP-02, equipped with
the laser diode head SLDH-041. The model has a peak power of $0.39\: mW$,
the pulse width (FWHM) is less than $30\: ps$, and the emitted light
of $410\: nm$ wavelength is close to the maximum sensitivity of the
ETL 9351 photocathode. The pulsing voltage supply of the laser provides
the internal trigger, which has a negligible time jitter ($<100\: ps$)
with respect to the light pulse. The laser head is mounted on a standard
optical bench together with the other elements, and is enclosed in
a light tight plastic box painted black inside (see Fig.\ref{Fig:OpticalSystem}).
The cable feed-throughs are light-tight as well, which is preventing
the pick up of the ambient light by the optic fiber. 

The light spot from the laser is intercepted by a supporting structure
designed to hold a set of neutral density filters that can easily
be changed or removed. An optical lens has been placed after the filters
to focus and reshape the laser spot; it can be aligned in three independent
directions by means of the micrometric regulators. The laser spot
is focused on the clad road, the support of the latter can be regulated
in two independent directions. The optical device has a high acceptance
and is especially designed to launch the pulse in a glass fiber of
600 $\mu$m core and 2 m length. The glass fiber is terminated by
an ST connector which is coupled in the same way to a bunch of optical
fibers, of 100 $\mu$m core and 2 m length that fit inside the glass
fiber spot. An optical grease is used to reduce the transmission losses
and reflections. All 12 outputs are terminated through ST patch panel
connectors that go out of the black box to deliver light into the
dark-room, to the WTs and to the TLTT. 

As regards the dark room, there are 4 fibers (12 m long) that, starting
from the black box, arrive over the 4 tables in the dark room. Each
fiber of the dark room and WT is supplied with a diffusing beam probe
with opal diffuser in order to provide a more uniform illumination
of the table, located at the minimum distance of 1.5 m from the fiber
termination. The diffuser has been especially studied and designed.

In view of the measurements of the relative PMT sensitivity the tables
were calibrated with a probe PMT, placed in turn at every position.
The illumination of the tables follows approximately the $1/r^{2}$
law, where $r$ is the distance between the PMT place and the diffuser.
The sensitivity of each PMT was defined during the tests with respect
to the calibration measurement. The values obtained should be considered
as indicative and have no direct use for the detector description
for a number of reasons, namely:

\begin{itemize}
\item the PMTs are sensitive to the residual EMF, but the orientation of
the PMT in respect to the probe was randomly chosen;
\item a different scheme of the EMF compensation was chosen for Borexino,
where the PMTs are screened by means of a $\mu$-metal;
\item because of the differences in the amplitude spectra, the same CFD
setting can leave different part of the signals under the threshold.
Moreover, the electronics of Borexino will be different from those
used in the test facility, which makes it impossible to adjust precisely
the thresholds.
\end{itemize}
The relative sensitivity of the PMTs can be easily defined during
the operation of the detector using events uniformly distributed over
the detector's volume, i.g. $^{14}C$ $\beta$-decay events.

\begin{figure*}
\begin{center}\includegraphics[%
  width=0.80\textwidth,
  height=0.35\textwidth]{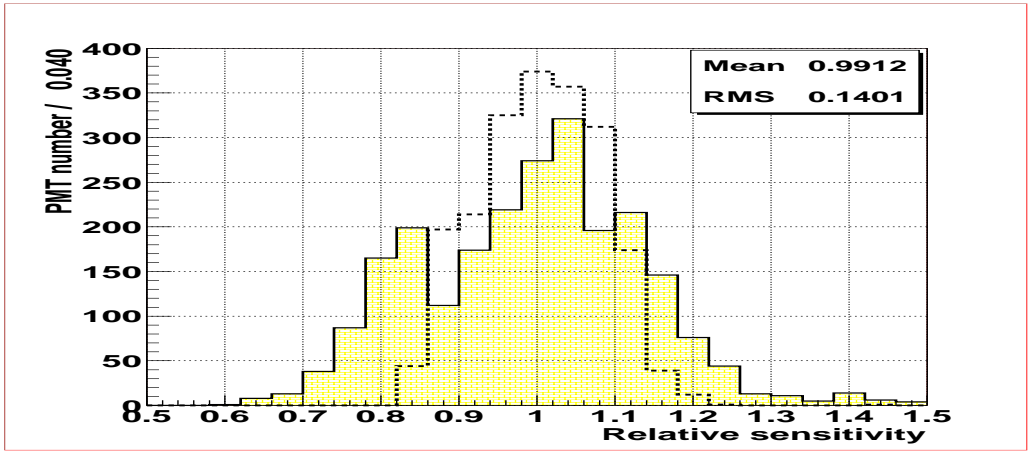}\end{center}

\caption{\label{Figure:RelSens}The relative sensitivity of the PMTs measured
in the test facility (filled histogram) in comparison with the ETL
provided measurements of the photocathode quantum efficiency at 420
nm.}
\end{figure*}

The relative sensitivity of the PMTs measured in the test facility
is shown in Fig. \ref{Figure:RelSens} in comparison with the ETL
provided measurements of the photocathode quantum efficiency at 420
nm (the values for each histogram are divided by their mean in order
to make comparison evident). The measured relative sensitivity has
bigger variations with respect to those of the quantum efficiency.

\subsection{Test procedure}

Before the installation in the test facility the PMTs were labelled.
60 PMTs were tested in one cycle of operation. The PMTs were installed
in the test facility in random positions on the test tables and with
random orientations of the dynodes plane. No special measures were
undertaken in order to prevent the illumination of the photocathodes
before settling them on the test tables. After leaving the PMTs in
the dark for a couple of hours, the nominal voltage provided by the
manufacturer was applied to the PMTs and the dark rate checked. In
the case of a PMT with a too high dark rate, the voltage has been
decreased in order to ensure stable dark count rate at the level of
less than 50 Kcps, and in addition the PMT was left under dark conditions
for a longer period of time (up to one week).

At the beginning of the test the expected current was calculated using
the known resistance of the high voltage divider and the operating
high voltage. If the deviation from the calculated value did not fall
in the fiducial interval, the PMT was examined more thoroughly. After
these preparative stages, the high voltage adjustment started as described
below (section \ref{Section:HVadjustment}). After the HV tuning finished,
the PMTs dark rates and currents were checked once more. The PMTs
still showing any kind of problem were excluded from the current test,
left under lower HV and reexamined after a couple of days.

\subsection{Software}

A special data acquisition (DAQ) and analysis software have been developed
for the test system. The kernel of the DAQ program was developed in
1992-1993 using Borland Turbo-C 3.0 under the MS-DOS operating system
with a computer on the base of an Intel 486 processor. Following the
rapid development of the computers, the program was adjusted to match
more powerful processors, and in its present state the data acquisition
speed is limited only by the hardware used. Now the program is running
in the MS-DOS emulator under a Linux system. 

The program provides control over all CAMAC settings of the electronics
(delays, thresholds, signals width etc) and high voltage supplies.
The program allows to specify the time period to save histograms on
disk, the time period to check the pedestal positions, the time period
to check PMT currents and high voltages, and the time period to check
the dark rate. The period of data taking was defined either by setting
the maximum number of events to be acquired, or by setting the run
time. The program provides also the possibility of the automatic switching
to the next group of PMTs, after the acquisition with the current
group is finished. 

The ADC pedestals are defined at the beginning of each specified period
and, if a shift of the position is found, the appropriate correction
is applied to the measured charges. In this way the low-frequency
drift of the pedestal (mainly due to the small daily temperature variations)
is compensated. 

The PMT currents are checked every specified time period and, if the
current is too high, the corresponding PMT is switched off automatically. 

In the {}``pmt acceptance test'' the data is acquired in the form
of 3 histograms of 1024 bins each: the charge spectrum (ADC), the
time of arrival of the response (TDC)%
\footnote{The full scale of TDC was set to 200 ns with 4096 channel resolution.
Because of the memory restrictions of the software, only part of the
full range is mapped to the histogram, specifically 100 ns in the
region {[}-30 ns;+70 ns{]} around the main peak in the PMT transit
time, with 1024 channel resolution. %
}, and the afterpulse spectrum up to 32 $\mu s$ (MTDC). The mean acquisition
rate is about 1000 events per second with 32 channels. 

The DAQ program can be run in a {}``pmt gain tuning'' mode. The
operation in this mode is described in more detail in the following
section. 

In the {}``dst'' mode of operation the data is written in a file
after each event, allowing to study correlations in the data. This
mode has been used in the optic fibers test \cite{Fibers}.

The analysis software was developed on the base of the CERN ROOT libraries
under a Linux system \cite{ROOT}. The program automatically analyses
the charge spectrum, the transit time spectrum, and the spectrum of
the ionic afterpulses and then plots all the data in the test sheet
(see Fig.\ref{fig:example} for an example). All numerical data is
inserted in a database immediately.

\section{\label{Section:HVadjustment}Electron multiplier gain measurements
and operating voltage setting}

\subsection{Brief description of the algorithm used}

For the HV tuning it is necessary to provide a robust method of gain
measurement. A method of photomultiplier calibration with high precision
of up to a few percent has been discussed in \cite{Filters}. The
method is based on precision measurements of the PMT charge response
to low intensity light pulses from a laser. It has been concluded
that the precision of the method is limited only by the systematic
errors in the discrimination of the small amplitude pulses from the
electronics noise. On the basis of our experience with the precise
PMT calibration \cite{Filters}, a fast procedure of PMT voltage tuning
has been realized for the Borexino PMT test facility \cite{HV}.

The goal of the tuning is to find the HV value that will provide an
electron gain factor $k=2\cdot10^{7}$ for each PMT. The mean value
of the charge single electron response (SER), $\: q_{1}$, is determined,
and the HV is adjusted so that $q_{1}$ agrees with a calibration
value $c_{1}=1.6\: pC$ to a predefined precision. In \cite{HV} was
shown that for small $\mu$ one can use the following approximation
for $q_{1}$:\begin{equation}
q_{1}=q_{m}\frac{1-\frac{\mu}{2}-p_{t}}{1-p_{t}\frac{thr}{2}},\label{eq:simplified}\end{equation}

where $q_{m}$ is the mean value of the cut distribution (a software
cut of $15\%$ of $c_{1}$ is used in order to avoid the effects of
the SER shape distortion near the hardware threshold);

$\mu$ is the mean p.e. number registered for one laser pulse;

$p_{t}$ is the part of the charge SER under the threshold. For the
15\% software threshold, the value $p_{t}\cong0.11$ was used, defined
during the tests of the 100 PMTs for the CTF; 

$thr$ is the threshold level measured in units of $c_{1}$(0.15 in
the our case).

The mean p.e. number is defined during the test by estimating the
probability of two sequential non-zero signals on PMT. Assuming a
Poisson distribution of the light detection process, one can write:

\begin{equation}
\mu=-\ln(1-\frac{N_{ev}}{N_{Triggers}})\simeq\frac{N_{ev}}{N_{Triggers}},\label{Eq:mu}\end{equation}

where $N_{Trigger}$ is the the full number of the system triggers
and $N_{ev}$ is the number of events that are followed by the non-zero
pulse (the first signal in a two pulses sequence is triggering the
system and thus is always present). In practice, we take as $N_{Trigger}$
the number of the events in the charge histogram (i.e. with a $5\%-10\%$
hardware cut), and as $N_{ev}$ we take the number of events after
a 30$\,$$\mu s$ delay%
\footnote{30$\,\mu s$ corresponds to the laser repetition rate of 33 kHz. The
laser repetition rate is fixed in the measurements at this value.%
} estimated from the Multihit TDC histogram (see Fig.\ref{fig:example}).
For these events the hardware cut on the CFD is about $20\%$ of the
SER. The precision of the $\mu$ estimation using these $N_{Triggers}$
and $N_{ev}$ values is approximately $10\%$, which is good enough
for our purpose.

The relative variation of the gain versus the variation of the applied
voltage was defined in \cite{HV}, performing the measurements of
the gain changes with applied voltages for 4 different PMTs:\begin{equation}
\frac{dk}{k}=(11-n)\cdot\frac{dU}{U-U_{D_{1}}},\label{formula:dk}\end{equation}

with a constant factor $n\simeq3.5$. $U_{D_{1}}=600$ V is the voltage
difference between the first dynode and the photocathode, the stability
of $U_{D_{1}}$ is provided by 3 Zener diodes of 200 V each (see Fig.\ref{fig:divider}).

The HV correction in the automated system is calculated in the following
way. If the deviation is bigger than $10\%$ the correction is set
to a fraction of the maximum deviation of 100$\:$V in proportion
: $(\frac{q_{1}-c_{1}}{c_{1}})$ to the deviation from the calibration
value. The value of 100$\,$V is used, so that any possible over-voltage
is avoided. For smaller deviations from the calibration the correction
can be calculated more precisely from (\ref{formula:dk}), namely: 

\begin{equation}
\Delta U=-\frac{3}{2}\frac{U-U_{D_{1}}}{11}\frac{q_{1}-c_{1}}{c_{1}}.\label{Eq:Delta_HV_correct}\end{equation}

\subsection{Results of operating voltage tuning}

\begin{figure*}
\begin{center}\includegraphics[%
  width=0.65\textwidth,
  height=0.45\textwidth]{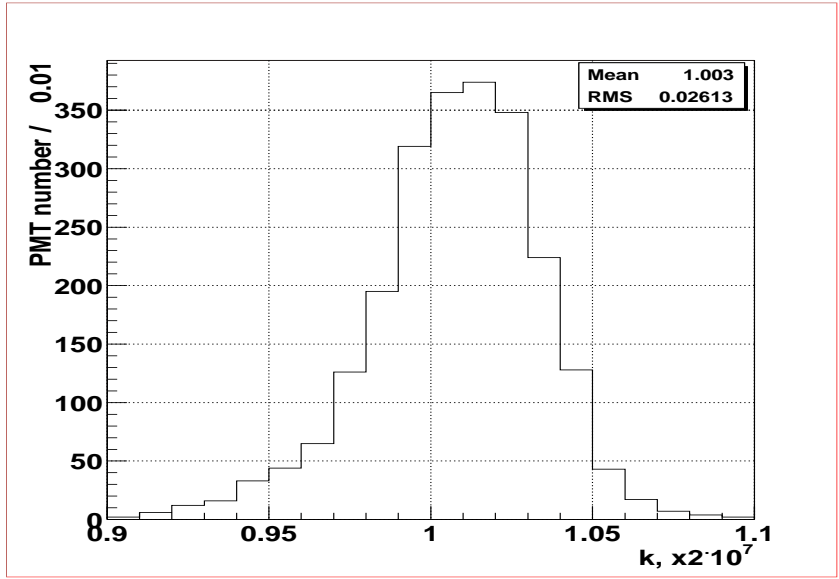}\end{center}

\caption{\label{fig:hv2000}Results of the PMT gain tuning. }
\end{figure*}

The algorithm is sufficiently fast; for 30$\:$PMTs at 1$\:$Kcps
acquisition rate the HV is adjusted in 15-20 minutes with $2\%$ statistical
precision.

\begin{figure*}
\begin{center}\includegraphics[%
  width=0.70\textwidth,
  height=0.45\textwidth]{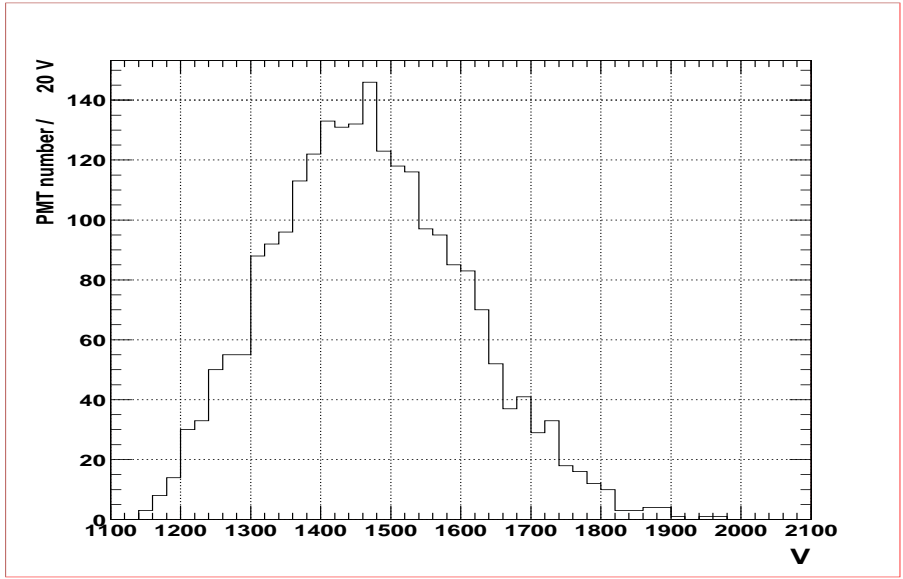}\end{center}

\caption{\label{Figure:HVstat}Operating high voltage for 2000 PMTs}
\end{figure*}

The results of the HV tuning with the precision set to $2\%\:$$\:$are
presented in Fig.\ref{fig:hv2000}. These results were obtained during
the high precision tests after the HV adjustment. The mean value of
k agrees with the expected $k=2.0\times10^{7}$. Statistics for the
operating voltage is shown in Fig.\ref{Figure:HVstat}.

The systematic error of the method is connected mainly with the substitution
of the parameter $p_{t}$ by its mean value $p_{t}=0.11$, the error
caused by the precision of $\mu$ estimation is negligible. The measured
variance of the $p_{t}$ parameter is $\sigma_{p_{t}}=0.018$ \cite{HV}.
As follows from the formula (\ref{eq:simplified}) the contribution
of the systematic error is equal to $\sigma_{p_{t}}$. Summing quadratically
the systematic (0.018) and the statistical (0.02) errors one will
obtain the full error of $\sigma\simeq0.027$. The variance of the
gain distribution (0.028) agrees with the calculated value. 

The method is not based on any specific single photoelectron spectrum
model and hence can be used for the gain adjustment of the PMT with
an arbitrary single electron response. Other advantages of the method
are the simplicity of calculations, the predictable precision and
very high speed of the algorithm. The method can be adapted for use
with any type of PMT designed to operate in single electron regime.

\section{Database}

The PostgreSQL \cite{PostreSQL} database was prepared and filled
with the data obtained during the tests. The data provided by the
manufacturer (ETL) were put in another table of the same database,
giving the possibility of easy access to the data. the complete description
of all database entries is described in \cite{Database}. The main
parameters put in the database are presented in Tables \ref{table: PMTQdata}-\ref{table:TDCandMTDC}.
These are mainly the parameters obtained directly in the test or after
the processing of the data. For the description of the PMT charge
spectrum the phenomenological model of the PMT charge spectrum was
used as described in \cite{Filters}, an example of the fit is presented
in Fig.\ref{fig:example}. Complete information on the measured parameters
will be provided elsewhere \cite{PMTsTest}.

\begin{table}

\caption{\label{table: PMTdata}PMT data}

\begin{center}\begin{tabular}{|c|l|}
\hline 
&
\multicolumn{1}{c|}{ Brief description of the database field}\tabularnewline
\hline
HV&
PMT Voltage corresponding approximately to the k=$2\times10^{7}$electronic
gain \tabularnewline
\hline
k&
\multicolumn{1}{l|}{Precise value of the PMT electron gain (in $2\cdot10^{7}$units)}\tabularnewline
\hline
$f_{dark}$&
Dark rate of the PMT in Kcps defined during the test. The threshold
for \tabularnewline
&
the counter is the same as for the TDC (see \textbf{tdcth})\tabularnewline
\hline
$S$&
Stability of the dark rate (variance of the observed dark rate divided
by the \tabularnewline
&
square root of the dark rate, for the stable dark rate should be close
to 1)\tabularnewline
\hline
$rs$&
Relative PMT sensitivity\tabularnewline
\hline
\end{tabular}\end{center}
\end{table}
\begin{table}

\caption{\label{table: PMTQdata}PMT Charge Spectrum data}

\begin{center}\begin{tabular}{|c|l|}
\hline 
&
\multicolumn{1}{c|}{ Brief description of the database field}\tabularnewline
\hline
$q_{1}$&
Mean value of the SER charge spectrum (pedestal is subtracted)\tabularnewline
&
estimated from the data\tabularnewline
\hline
p/v&
Peak to Valley ratio\tabularnewline
\hline
$\mu$&
Mean value of photoelectrons per laser trigger defined from the \tabularnewline
&
Charge Spectrum Histogram as $\mu=-ln(\frac{N_{ped}}{N_{Trigger}})$\tabularnewline
\hline
$rs$&
Relative PMT sensitivity\tabularnewline
\hline
&
\multicolumn{1}{l|}{Relative variance of the Single Electron Response charge Spectrum }\tabularnewline
$v_{1}$&
(i.e. $\frac{\sigma_{q_{1}}^{2}}{q_{1}^{2}}$, where $\sigma_{q_{1}}$and
$q_{1}$are the rms and mean value of the SER\tabularnewline
&
charge Spectrum) estimated for the data\tabularnewline
\hline
$p_{t}$&
Part of the SER charge spectrum under the TDC (hardware) threshold
\textbf{tdcth}\tabularnewline
&
(estimated without fit)\tabularnewline
\hline
\end{tabular}\end{center}
\end{table}

\begin{table}

\caption{\label{table:PMTfit}The parameters of the model of single electron
charge spectrum}

\begin{center}\begin{tabular}{|c|l|}
\hline 
&
\multicolumn{1}{c|}{ Brief description of the database field}\tabularnewline
\cline{1-1} 
\hline 
$\mu$&
Mean value of photoelectrons per laser trigger defined from the fit.\tabularnewline
\hline
$q_{1}$&
Histogram mean value of the SER charge Spectrum (estimated from the
fit)\tabularnewline
\hline
$v_{1}$&
Relative variance of the Single Electron Response charge Spectrum \tabularnewline
&
calculated from fit values. \tabularnewline
\hline
$q_{0}$&
Mean of the Gaussian part of SER charge spectrum\tabularnewline
\hline
$s_{0}$&
RMS of the Gaussian part of SER charge spectrum \tabularnewline
\hline
$p_{U}$&
Fraction of the the events under the exponential part of SER (underamplified)\tabularnewline
\hline
$A$&
Slope of the exponential part of SER \tabularnewline
\hline
\end{tabular}\end{center}
\end{table}

\begin{table}

\caption{\label{table:TDCandMTDC}PMT transit time characteristics}

\begin{center}\begin{tabular}{|c|l|}
\hline 
parameter&
\multicolumn{1}{c|}{Brief description of the database field}\tabularnewline
\hline
\textbf{tdcth}&
Threshold of the CFD calculated from the events number in the TDC\tabularnewline
&
histogram (in p.e.)\tabularnewline
\hline 
$t_{0}$&
position of the Gaussian peak on the TDC histogram ({[}ns{]})\tabularnewline
\hline
$\sigma_{t}$&
rms of the Gaussian peak on the TDC histogram ({[}ns{]})\tabularnewline
\hline
$p_{late}$&
late pulsing in percents (estimated as the ratio of the events in\tabularnewline
&
{[}t0+3tsig;100 ns{]} range to the events number under the Gaussian
peak\foreignlanguage{english}{)}\tabularnewline
\hline
$p_{early}$&
early pulses in percents (estimated as the ratio of the events in\tabularnewline
&
{[}0;t0-3tsig ns{]} range to the events number under the Gaussian
peak)\tabularnewline
\hline
\end{tabular}\end{center}
\end{table}
\begin{table}

\caption{\label{table:MTDC}Afterpulses characteristics}

\begin{center}\begin{tabular}{|c|l|}
\hline 
parameter&
\multicolumn{1}{c|}{Brief description of the database field}\tabularnewline
\hline
$a_{1}$&
percentage of afterpulses in the {[}0.4-1.0{]} $\mu s$ range\tabularnewline
\hline
$a_{2}$&
percentage of afterpulses in the {[}1.0-3.6{]} $\mu s$ range\tabularnewline
\hline
$a_{3}$&
percentage of afterpulses in the {[}3.6-12.4{]} $\mu s$ range\tabularnewline
\hline
$a_{tot}$&
percentage of afterpulses in the {[}0.4-12.4{]} $\mu s$ range\tabularnewline
\hline
\end{tabular}\end{center}
\end{table}

\section{Concluding remarks}

The main task of the test system was the acceptance test of the Borexino
PMTs. The system proved to be very efficient, 2350 PMTs delivered
from the manufacturer were tested in 4 months of continuous system
operation. The results of the test were analyzed and will be reported
elsewhere \cite{PMTsTest}. Electronics and software of the test system
have been used also in the Borexino Muon Veto PMTs test; in Borexino
light fibers tests \cite{Fibers}; in testing of 120 PMTs for the
upgrade of the CTF detector; in the charge spectrum study of the PMT
\cite{Filters}; in time characteristics high precision study of ETL9351
type PMT \cite{TDC}; and in the study of the sensitivity of the PMT
to the Earth magnetic field \cite{EMFcompensation}.

\section{Acknowledgements}

We are deeply grateful to G.Bacchiocchi, R.Cavaletti, R.Dossi, D.Giugni,
P.Saggese, S.Grabar, and R.Scardaoni who were involved in various
activities during the test facility installation, and in particular
to G.Korga and L.Papp who took an active part in the test. We also
thank the LNGS staff for the warm atmosphere and the good working
conditions. The job of one of us (O.S.) was supported by the INFN
sez. di Milano, and he is personally indebted to Prof. G.Bellini for
the possibility to work at the LNGS laboratory. The authors appreciate
the help of M.Laubenstein in preparation of the manuscript.

Credit is given to the developers of the CERN ROOT program \cite{ROOT},
that was used in the calculations and to create all the figures of
the article; as well as to the developers of the PostgreSQL database
\cite{PostreSQL}.

\end{document}